\documentclass[preprint,12pt,dvipdfmx]{elsarticle}

\usepackage{graphicx}

\usepackage{amssymb}
\usepackage{latexsym}

\journal{Physica A}

\begin{document}

\begin{frontmatter}



\title{
Criticality of
the Higgs mass 
for the long-range quantum $XY$ chain:
Amplitude ratio between the Higgs and paramagnetic gaps
}


\author{Yoshihiro Nishiyama} 

\address{Department of Physics, Faculty of Science,
Okayama University, Okayama 700-8530, Japan}

\begin{abstract}
The quantum $XY$ spin chain
with 
the interactions decaying as a power law $1/r^{1+ \sigma}$
of the distance between spins $r$
was investigated with the exact diagonalization method.
Here, the constituent spin
is set to $S=1$,
which enables us to incorporate the
biquadratic interactions
so as to realize
the order-disorder transition with the O$(2)$ symmetry maintained.
Thereby, in the ordered phase,
we resolved the Higgs mass $m_H$
out of the Goldstone-excitation continuum
by specifying 
Higgs-particle's
quantum numbers to adequate indices.
We then turn to the analysis of the critical amplitude ratio
$m_H / \Delta$ ($\Delta$: paramagnetic gap in the disordered phase).
As the power of the algebraic decay $\sigma$ increases,
the amplitude ratio $m_H/\Delta$ gets enhanced gradually
in agreement with 
the $\epsilon(=4-D)$-expansion-renormalization-group result;
here, we resort to the $\sigma \leftrightarrow D$
relation advocated recently in order to establish a relationship
between the renormalization-group result and ours.

\end{abstract}


\end{frontmatter}



\section{\label{section1}Introduction}

\begin{figure}
\includegraphics[width=120mm]{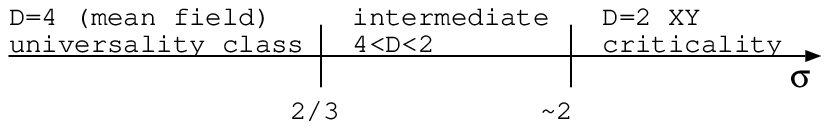}%
\caption{
\label{figure1}
The universality chart 
\cite{Defenu17,Campana10,Gong16,Maghrebi17,Frerot17,Roy18}
for
the quantum $XY$ spin chain 
with the interactions decaying as a power law 
$1/r^{1+\sigma}$ of the distance between spins $r$
is presented.
For $\sigma  \gtrapprox 2$, the long-range interaction becomes irrelevant,
and the universality class of the order-disorder phase transition
falls into that of the two-dimensional ($D=2$) classical $XY$ model,
whereas for $\sigma > 2/3$,
the system exhibits the mean-field ($D=4$) criticality.
In the intermediate regime $2/3 \le \sigma \lessapprox 2$,
the criticality with fractional effective dimensionality
$2 \le D \le 4$ is realized.
That is, the power of the algebraic decay $\sigma$
interpolates smoothly the upper and lower critical dimensions.
At these boundaries, there arise logarithmic corrections
\cite{Luijten02,Brezin14,Defenu15,Fey16}
to finite-size scaling.
}
\end{figure}

The
spin systems with the long-range interactions
have been investigated
both
theoretically
\cite{Fisher72,Sak73,Luijten02,%
Picco07,Blanchard13,Grassberger13,Gori17,Angelini14,%
Joyce66,Brezin14,Defenu15,Defenu16,Goll18,%
FloresSola17,Horita17,Sun96}
and experimentally
\cite{Wu91,Britton12,Islam13,Richerme14,Jurcevic14,Paz13}.
A main concern is to clarify how the criticality
between the ordered and disordered phases is affected by
the power of the algebraic decay
\cite{Gori17,Angelini14,Joyce66}.
Meanwhile, the idea was extended to the case of the 
{\em quantum} spin models
\cite{Laflorencie05,Dutta01,Defenu17,Fey16,Sandvik10,Koffel12,Humeniuk16,%
Campana10,Gong16,Maghrebi17,Frerot17,Roy18}.
In Fig. \ref{figure1},
we present the criticality chart
\cite{Defenu17,Campana10,Gong16,Maghrebi17,Frerot17,Roy18}
for the 
quantum $XY$ spin chain with the interactions
decaying as a power law 
$1/r^{1+\sigma}$ of the distance between spins $r$.
For $\sigma \gtrapprox 2$, the long-range interaction becomes irrelevant,
and the criticality falls into the universality class
of the classical two-dimensional ($D=2$)
$XY$ model (with the short-range interactions).
On the contrary,
for $\sigma < 2/3$,
the criticality belongs to
the mean-field type, namely, 
the $D=4$ universality class.
In the intermediate regime $2/3 \le \sigma \lessapprox 2 $,
the power $\sigma$
interpolates both limiting cases, $D=2$ and $4$,
smoothly.
Actually, relying on the
correspondence between the $\sigma$- and
$D$-mediated considerations,  
the authors of Ref. \cite{Goll18}
analyzed
the criticality for the $D=3$ (short-range) Ising model
quantitatively.
Experimentally,
 systems with variable-range interactions
have been synthesized
\cite{Britton12,Islam13,Richerme14}.
In this sense,
the $\sigma \leftrightarrow D$ relation \cite{Angelini14}
is not a mere theoretical concept.

\begin{figure}
\includegraphics[width=120mm]{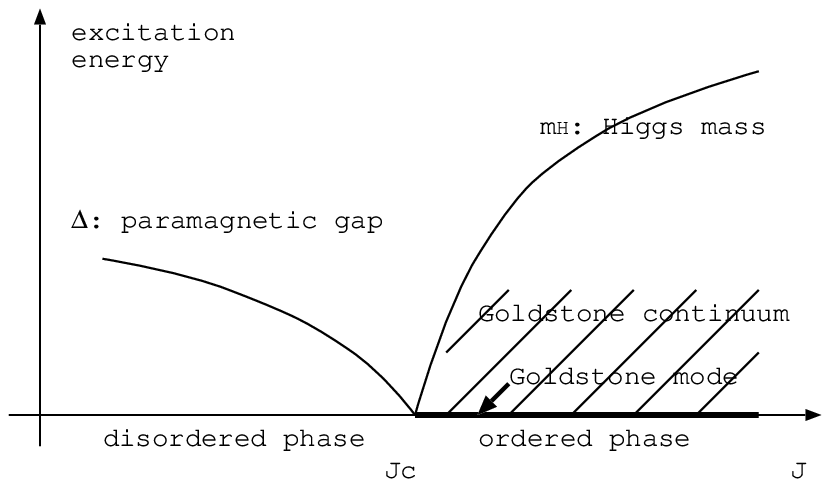}%
\caption{
\label{figure2}
A schematic drawing of the low-lying spectrum for
the quantum long-range $XY$ spin model 
(\ref{Hamiltonian})
with the $XY$ interaction $J$ 
is presented.
In the ordered phase $J>J_c$,
there opens
the Higgs gap 
$m_H$
embedded within the Goldstone-excitation continuum,
whereas in the disordered phase 
$J<J_c$,
the paramagnetic gap
$\Delta$ appears.
The scaling behavior for $m_H$ as well as
the critical amplitude ratio 
$m_H(J)/\Delta(2J_c-J)|_{J\to J_c^+}$ are the main concern.
}
\end{figure}

The criticality chart, Fig. \ref{figure1},
resembles to that of 
the classical counterpart \cite{Fisher72,Sak73,Gori17}.
In this paper,
aiming to elucidate the 
{\em quantum} nature of this problem,
we devote ourselves to
the spectrum, namely, the Higgs (paramagnetic) excitation gap,
$m_H$ ($\Delta$), in the (dis)ordered phase.
Particularly,
we estimate
the critical amplitude
ratio 
\begin{equation}
\label{amplitude_ratio}
m_H/\Delta=\left. \frac{m_H(J)}{\Delta (2J_c-J)} \right|_{J\to J_c^+}
,
\end{equation}
with the $XY$ interaction $J$ and the power $\sigma$ ranging
within $2/3 < \sigma<2$.
In Fig. \ref{figure2},
we present a schematic drawing 
for the spectrum in the
disordered ($J<J_c$)
and ordered 
($J>J_c$) phases.
The paramagnetic gap opens in 
the disordered phase,
whereas
in the ordered phase, 
the Higgs branch $m_H$
is embedded within the
 Goldstone-excitation continuum.
Meanwhile,
the amplitude ratio 
$m_H/\Delta$ was estimated
for the generic values of $D$
by means of the $\epsilon$-expansion method \cite{Katan15}.
For the fixed-$D=3$ systems, a good deal of analyses
have been made
\cite{Gazit13a,Gazit13b,Chen13,Nishiyama15,Rancon14,Rose15}.
As mentioned above,
via the $\sigma \leftrightarrow D$ relation
\cite{Angelini14,Defenu17},
one is able to establish a relationship among
these $D$- and $\sigma$-mediated considerations.

We employed the exact diagonalization method,
which enables us to calculate
the energy gaps, $m_H$ and $\Delta$,
directly
by
specifying the quantum numbers 
to
Eqs. (\ref{Higgs_gap}) and 
(\ref{paramagnetic_gap}), respectively.
So far, in the quantum-Monte-Carlo approach
\cite{Gazit13a,Gazit13b,Chen13},
the dynamical scalar susceptibility
$\chi_s(\omega)$
\cite{Dupuis11,Podolsky11}
has been enumerated through the inverse Laplace
transformation;
see Appendix B of Ref. \cite{Gazit13b}.
Because the disturbance $\chi_s(\omega)$ preserves the O$(2)$ symmetry,
it is less sensitive to the Goldstone excitations
\cite{Pekker15};
hence, a $m_H$ signal can be captured
via $\chi_s(\omega)$ \cite{Podolsky11}.
In the recent experiment for the ultra cold atom \cite{Endres12},
essentially the same strategy is undertaken;
for such boson system,
the (dis)ordered phase should be interpreted as
the superfluid (Mott insulator) phase.
We stress that the exact diagonalization method admits $m_H$ and $\Delta$
without resorting to such elaborated techniques.

To be specific,
we present 
the Hamiltonian for the one-dimensional
quantum spin-$S=1$ $XY$ model with the long-range interactions
\begin{equation}
\label{Hamiltonian}
{\cal H}
=- \frac{1}{\cal N} \sum_{i \ne j} J_{ij}
  \left(\frac{J}{2}(S^+_i S^-_j + S^-_i S^+_j)  
 + \frac{J_4}{4}((S^+_i)^2(S^-_j)^2+(S^-_i)^2(S^+_j)^2)
   \right)
+ D_s \sum_{i=1}^{N} (S^z_i)^2
    .
\end{equation}
Here, the quantum spin-$S=1$ operator 
$S^{\pm,z}_i$
is placed at each one-dimensional-lattice point, $i=1,2,\dots,N$.
The summation 
$\sum_{i \ne j}$
runs over all possible pairs $1 \le i,j \le N$,
and the long-range
interaction $J_{ij}$
decays as a power law,
$J_{ij} = 1/ \sin (\pi |i-j| /N)^{1+\sigma}$,
with the variable exponent $\sigma$.  
Here,
we implemented 
the periodic-boundary condition for the spin chain;
what is meant by
the expression for $J_{ij}$
is
that 
the distance between the spins, $i$ and $j$,
is given by
the chord length, $\sin(\pi |i-j| /N)$,
rather than the arc, $|i-j|$.
The $XY$ interaction $J$ induces the order-disorder transition (see Fig. \ref{figure2}),
whereas the biquadratic interaction $J_4=1$ 
\cite{Rousseau05,Sandvik02,Emidio16}
is fixed.
The single ion anisotropy $D_s=0.1(2-\sigma)$ 
is included so as to improve the finite-size behaviors
\cite{Deng03};
the single ion anisotropy drives the $XY$ phase
to the massive phase \cite{Botet83,Glaus84,Solyom84} in $D=2$ dimensions
($\sigma=2$),
and we incorporated the diminution factor $(2-\sigma)$.
The normalization (Kac) factor
${\cal N}$
\cite{Homrighausen17,Vanderstraeten18}
is given by
${\cal N}=N^{-1} \sum_{i \ne j} \sin(\pi |i-j|/N)^{-1-\sigma}$.

The rest of this paper is organized as follows.
In the next section,
we show the numerical results.
A brief account of the simulation algorithm
is given as well.
In the last section,
we address the summary and discussions.

\section{\label{section2}Numerical results}

In this section, we present the numerical results.
We employed the exact diagonalization method
for the Hamiltonian (\ref{Hamiltonian}) with $N \le 22$ spins.
Before commencing detailed scaling analyses for $m_H$ and $\Delta$,
we give a brief account of the numerical algorithm.
The Hamiltonian has a number of symmetries,
{\it i.e.}, good quantum numbers,
with which we are able to
diagonalize the Hamiltonian
within the restricted Hilbert space.
The quantum-number specification scheme for $m_H$ and $\Delta$ is as follows:
The numerical diagonalization was performed within the zero-momentum subspace
$k=0$, at which the elementary-excitation gap opens;
namely, both the ground- and first-excited-state levels locate.
We further specify the subspace with an additional quantum number $S^z_{tot}$,
which corresponds to the total magnetization operator, $\sum_{i=1}^N S^z_i$;
this quantum number reflects the O$(2)$ symmetry of ${\cal H}$.
Within the restricted Hilbert space specified by $k=0$ and $S^z_{tot}$,
the Higgs mass is given by
\begin{equation}
\label{Higgs_gap}
m_H=E_1(0)-E_0(0)
,
\end{equation}
with the ground- (first excited-) state energy
$E_{0(1)}(S^z_{tot})$ within the sector $S^z_{tot}$.
The paramagnetic excitation belongs to the $S^z_{tot}= \pm 1$ sector,
and hence, the gap is given by the formula
\begin{equation}
\label{paramagnetic_gap}
\Delta=E_0(1)-E_0(0)
.
\end{equation}
This $\Delta$ branch becomes
 the Goldstone mode in the adjacent phase (ordered phase).
Note that the $m_H$ mode exists far above the collection of the Goldstone excitations (continuum),
and such level specification is significant to resolve the former 
out of the latter.

\subsection{\label{section2_1}
Scaling analysis of the dynamical critical exponent $z$:
A preliminary survey}

In this section,
based on
the scaling theory developed in Ref. \cite{Defenu17},
we analyze the Higgs mass $m_H$, Eq. (\ref{Higgs_gap}).
As a byproduct,
we obtain the dynamical critical exponent $z$,
which characterizes the anisotropy
between the real-space and imaginary-time directions
quantitatively.

\begin{figure}
\includegraphics[width=120mm]{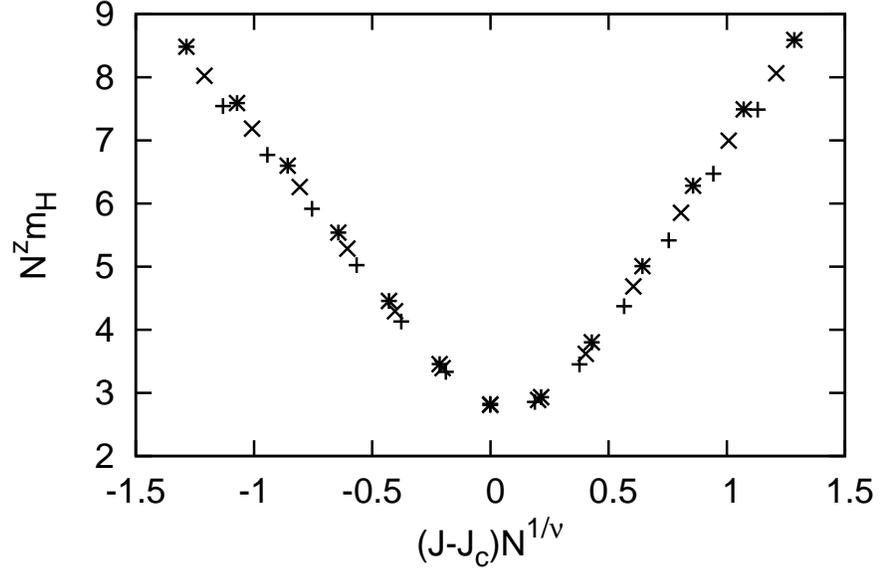}%
\caption{
\label{figure3}
The scaling plot,
$(J-J_c)N^{1/\nu}$-$N^z m_H$,
is presented for $\sigma=1.2$, and
the system sizes,
($+$) $N=18$,
($\times$) $20$,
and ($*$) $22$,
with 
the scaling parameters,
$J_c=0.30785$, 
$z=0.594$ and, 
$1/\nu=0.636$.
The Higgs gap opens in the ordered phase $J>J_c$.
This mode seems to be massive in the adjacent phase
$J<J_c$ as well,
supporting the preceding observations
with the 
quantum Monte Carlo method
\cite{Gazit13b,Chen13}.
}
\end{figure}

In Fig. \ref{figure3},
we present the scaling plot,
$(J-J_c)N^{1/\nu}$-$N^z m_H$,
for the fixed $\sigma=1.2$
and various system sizes,
($+$) $N=18$, ($\times$) $20$, and ($*$) $22$. 
The scaling parameters, namely,
the critical point $J_c=0.30785$,
the reciprocal correlation-length critical exponent 
$1/\nu=0.636$,
and the dynamical critical exponent $z=0.594$,
were determined as follows.
The critical point 
$J_c=0.30785(40)$ was obtained
through the least-squares fit for 
the 
$1/N$-$J_c(N)$ data
with $N=16,18,\dots,22$;
here,
the approximate critical point $J_c(N)$
is defined by
\begin{equation}
\partial_J m_H |_{J=J_c(N)}=0
,
\end{equation}
for each $N$.
Likewise, the dynamical critical exponent 
$z=0.594(1)$   
was obtained via the least-squares fit for
the
$\left( \frac{N+(N+2)}{2} \right)^{-1}$-$z(N,N+2)$ 
data
with $N=14,16,\dots,20$.
The approximate dynamical critical exponent
$z(N,N')$ is given by the logarithmic derivative of the 
finite-size Higgs mass
\begin{equation}
z(N,N')=-\frac{\ln m_H(N)|_{J_c(N)}-\ln m_H(N')|_{J_c(N')}}{\ln N -\ln N'}
 ,
\end{equation}
with a pair of system sizes $(N,N')$.
Finally,
the reciprocal correlation-length critical exponent
$1/ \nu =0.636$
was calculated from the above-mentioned result $z=0.594$
with the aide of $1/z\nu=1.07$
taken from Fig. 3a of Ref. \cite{Defenu17}.

The scaled data in Fig. \ref{figure3} seem to obey the finite-size scaling
satisfactorily.
The Higgs mass appears to open in the ordered phase $J>J_c$.
The criticality for $m_H$ is analyzed in depth afterwards.
Noticeably enough,
the $m_H$ branch opens also in the adjacent (disordered) phase $J<J_c$.
This massive mode may correspond to the particle-hole-excitation threshold
in the boson language \cite{Gazit13b,Chen13};
further details 
as to the disordered phase
are not pursued in this paper.

\begin{figure}
\includegraphics[width=120mm]{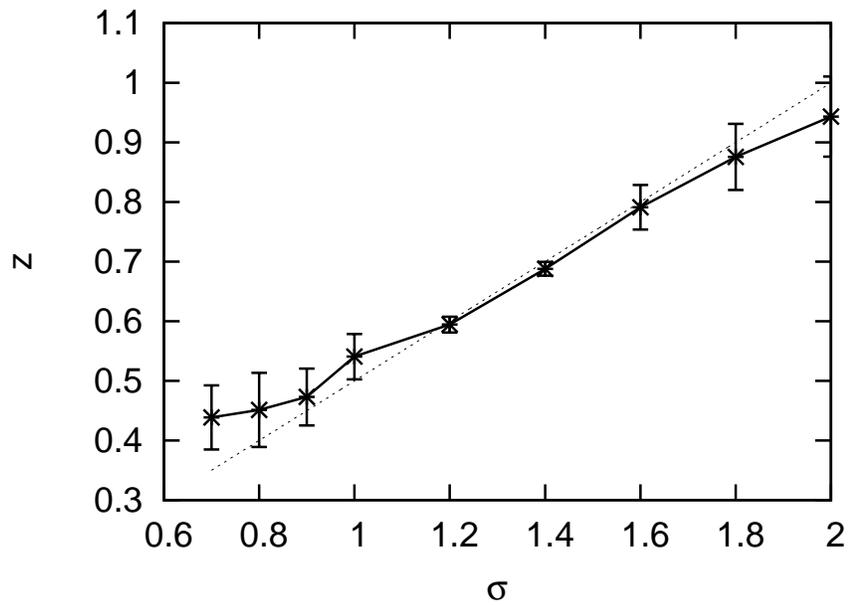}%
\caption{
\label{figure4}
The dynamical critical exponent 
$z$ is presented 
for the power of algebraic decay $\sigma$.
An approximate formula
\cite{Defenu17},
 Eq. (\ref{z_formula}),
is presented (dotted) as well.
This formula is validated at least for small $\sigma$.
Deviations around
the lower and upper critical thresholds,
$\sigma=2/3$ and $2$, respectively,
may be attributed to the notorious
logarithmic corrections to 
finite-size scaling 
\cite{Luijten02,Brezin14,Defenu15,Fey16}.
}
\end{figure}

Similar analyses as that of Fig. \ref{figure3}
were carried out for various values of $\sigma$.
The $\sigma$-dependent dynamical critical exponent
$z$ is presented in
 Fig. \ref{figure4}.
Here, as an error margin,
we accepted
the deviation between 
different extrapolation schemes,
namely,
$N^{-1}$-$z(N,N+2)$ and 
$N^{-2}$-$z(N,N+2)$ approaches,
aiming to appreciate possible systematic errors
other than the least-squares-fit error.
The series of results in Fig. \ref{figure4} indicate that
the dynamical critical exponent $z$ reflects the variation
of $\sigma$ sensitively.
As a reference, in Fig. \ref{figure4},
we also present an approximate formula
(dots)
\cite{Defenu17}
\begin{equation}
\label{z_formula}
z = \sigma /2,
\end{equation}
which is validated
in the small-$\sigma$ side.
Our data appear to obey the formula (\ref{z_formula})
for a considerably wide range of $\sigma$.
However, 
around the upper and lower critical thresholds,
$\sigma=2$ and $2/3$, respectively,
there emerge systematic deviations,
which may be attributed to the
notorious logarithmic corrections to finite-size scaling
\cite{Luijten02,Brezin14,Defenu15,Fey16}.
Note that these thresholds correspond
to the lower and upper critical dimensions,
$D=2$ and $4$, respectively (see Eq. (\ref{sigma-D_relation})),
and inevitably,
the finite-size scaling is affected by the marginal operators
at these thresholds.


We address a number of remarks.
First,
the dynamical critical exponent $z$ 
is peculiar to the 
 {\em quantum} criticality.
The conventional short-range spin models
exhibit $z=1$, namely,
the restoration of the symmetry between
the real-space and imaginary-time subspaces.
On the contrary,
as shown above,
the long-range counterpart does exhibit
the anisotropy,
$z \ne 1$ \cite{Defenu17}, between these subspaces.
In the quantum Monte Carlo simulation, a special care has to be paid
so as to remedy this anisotropy in order to manage the finite-size-scaling
properly.
In the exact diagonalization approach,
one is able to concentrate on the real-space sector, because
the imaginary-time
 system size is infinite $\beta(=1/T) \to \infty$ {\it a priori};
note that
one can access the ground state $T=0$ directly.
Recently, as for the quantum long-range systems,
a variety of approaches such as
the exact diagonalization method \cite{Homrighausen17},
the variational matrix product state with the quasi-particle ansatz \cite{Vanderstraeten18},
and
the density-matrix renormalization group under the {\it open} boundary condition \cite{Frerot18},
have been utilized successfully.
Because 
in the present analysis,
the mass gaps right at the zone center $k=0$ are required, 
we resort to the exact diagonalization method under the
 periodic boundary condition.
Last,
in
 Fig. \ref{figure3},
the minimum point for the Higgs gap seemingly locates
slightly out of
 the critical point $(J-J_c)N^{1/\nu} \approx 0.1$.
However,
in the thermodynamic limit,
the deviation from the critical point $J_c$ vanishes 
 as  $J-J_c \approx 0.1/N^{1/\nu} \to 0$.

\subsection{\label{section2_2}
Critical amplitude ratio $m_H/\Delta$}

\begin{figure}
\includegraphics[width=120mm]{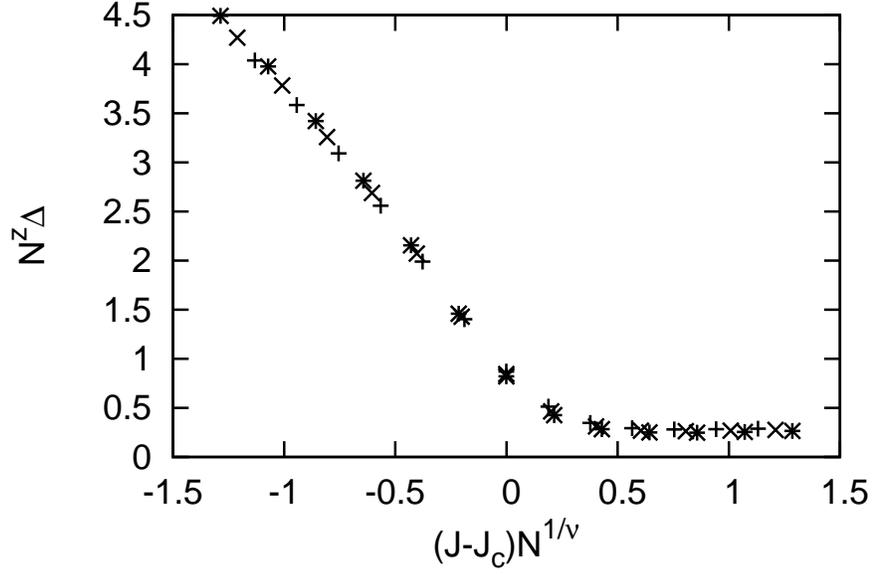}%
\caption{
\label{figure5}
The scaling plot,
$(J-J_c)N^{1/\nu}$-$N^z \Delta$,
is presented for $\sigma=1.2$, and
the system sizes,
($+$) $N=18$,
($\times$) $20$,
and ($*$) $22$.
The scaling parameters,
$J_c=0.30785$, 
$z=0.594$ and 
$1/\nu=0.636$, 
are the same as those of Fig. \ref{figure3}.
The paramagnetic gap $\Delta$
opens in the disordered phase $J < J_c$.
This branch becomes the Goldstone gapless mode in the adjacent phase
$J>J_c$, forming the Goldstone continuum in the low-energy spectrum.
}
\end{figure}

Based on the 
finite-size-scaling analysis examined in the preceding section,
we turn to the analysis of the critical amplitude ratio 
$m_H/\Delta$.
To this end,
we investigate the 
scaling behavior of the paramagnetic gap $\Delta$, Eq.
(\ref{paramagnetic_gap}).
In Fig. \ref{figure5}, we present the scaling plot,
$(J-J_c)N^{1/\nu}$-$N^z \Delta$,
for $\sigma=1.2$ and
various system sizes,
($+$) $N=18$,
($\times$) $20$,
and 
($*$) $22$.
The scaling parameters,
$J_c=0.30785$, 
$1/\nu=0.636$, and $z=0.594$,
are the same as those of Fig. \ref{figure3}.
The scaled data fall into the scaling curve satisfactorily.
The paramagnetic gap $\Delta$ opens in the disordered phase 
$J<J_c$; see the schematic diagram, Fig. \ref{figure2}, as well.
The paramagnetic mode $\Delta$ in the disordered phase
turns into the Goldstone excitation in the adjacent phase 
(ordered phase) $J>J_c$.
Actually,
in $J>J_c$,
there is shown a closure of the $\Delta$ branch,
which is smaller than the Higgs gap $m_H$ in Fig. \ref{figure3}.
Such Goldstone continuum place an obstacle to observe $m_H$ clearly.
In this paper, we pick up the $m_H$ mode
directly by specifying
the quantum numbers to Eq. (\ref{Higgs_gap}) adequately.

\begin{figure}
\includegraphics[width=120mm]{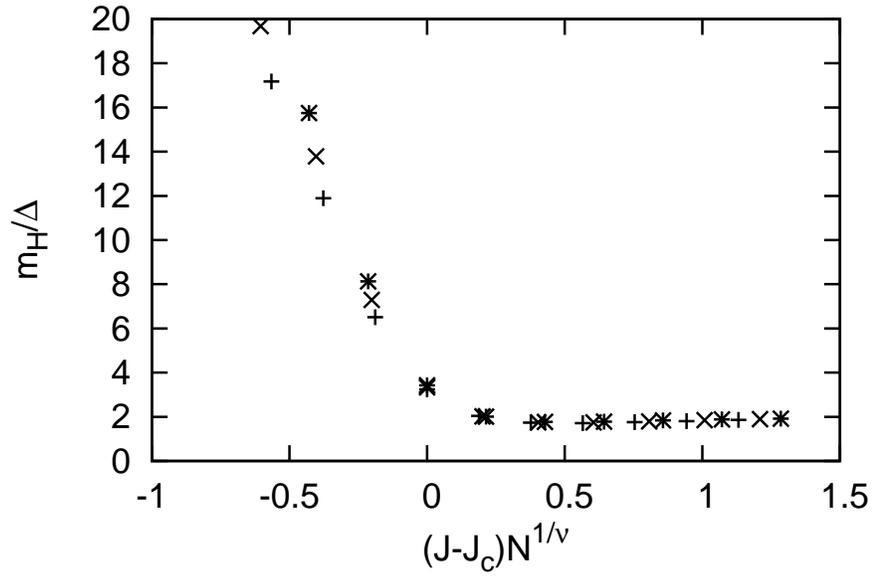}%
\caption{
\label{figure6}
The scaling plot,
$(J-J_c) N^{1/\nu}$-$m_H(J)/\Delta(2J_c-J)$,
is presented for $\sigma=1.2$, and
the system sizes,
($+$) $N=18$,
($\times$) $20$,
and ($*$) $22$.
The scaling parameters,
$J_c=0.30785$, and 
$1/\nu=0.636$, 
are the same as those of Fig. \ref{figure3}.
In the ordered phase $J > J_c$,
there appears a plateau 
with the height $m_H/\Delta \approx 1.8$, 
which admits an estimate for the amplitude ratio;
see text for details.
}
\end{figure}

We then
consider the amplitude ratio $m_H/\Delta$.
In Fig. \ref{figure6},
we present the scaling plot,
$(J-J_c)N^{1/\nu}$-$m_H(J)/\Delta(2J_c-J)$,
for $\sigma=1.2$ and 
various system sizes,
($+$) 
$N=18$,
($\times$)
$20$,
and
($*$) $22$.
The scaling parameters,
$J_c=0.30785$, and $1/\nu=0.636$,
are the
same as those of Fig. \ref{figure3}.
Again, the collapse of the scaled data seems to be satisfactory.
In the ordered phase 
$J>J_c$,
the amplitude ratio takes a plateau with the height $m_H/\Delta \approx 1.8$.
Such a feature indicates that
that the amplitude ratio $m_H/\Delta$ indeed takes a constant value
in proximity to the critical point.

The plateau in Fig. \ref{figure6}
exhibits a shallow bottom, which provides a good indicator
as to the amplitude ratio for each $N$.
Aiming to extrapolate $m_H/\Delta$ systematically to the thermodynamic 
limit $N \to \infty$,
we carried out
the least-squares fit for the plot
$N^{-1}$-$\frac{m_H}{\Delta}(N)$
with $N=16,18,\dots,22$;
here 
the approximate amplitude ratio
is given by
\begin{equation}
\frac{m_H}{\Delta}(N)=
\left.
\frac{m_H(J)}{\Delta(2J_c-J)} \right|_{J=\bar{J}(N)}
,
\end{equation}
at the bottom of the plateau $\bar{J}(N)$
satisfying the extremum condition
\begin{equation}
\left.
\frac{\partial \left( m_H(J)/\Delta(2J_c-J) \right)}
{\partial J}
\right|_{J=\bar{J}(N)} =0
,
\end{equation}
for each $N$.
We arrived at
$m_H/\Delta=1.980(14)$ via the
extrapolation scheme.

\begin{figure}
\includegraphics[width=120mm]{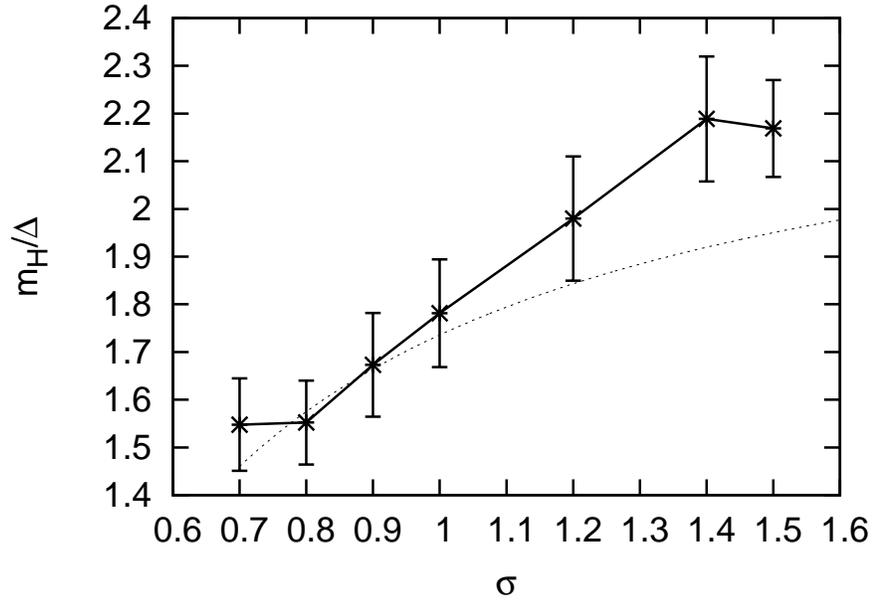}%
\caption{
\label{figure7}
The critical amplitude ratio $m_H/\Delta$,
Eq.
(\ref{amplitude_ratio}),
is presented for various $\sigma$.
The series of solutions terminate around
$\sigma \approx 1.6$,
because
the Higgs excitation becomes unstable for 
exceedingly large $\sigma \gtrapprox 1.6$.
According to the formula (\ref{sigma-D_relation}),
the power $\sigma \approx 1.6$ corresponds to
$D \approx 2.25$, which is about to reach the lower critical dimension $D=2$.
As a reference, the $\epsilon(=4-D)$-expansion formula
\cite{Katan15},
Eq. (\ref{peak_width}), is shown as a dotted curve.
}
\end{figure}

We carried out similar
analyses as that of Fig. \ref{figure6}
for various values of $\sigma$.
The $\sigma$-dependent
amplitude ratio $m_H/\Delta$
is presented in Fig. \ref{figure7}.
Here, as an error margin,
we accepted the deviation between different extrapolation
schemes,
namely,
$N^{-1}$-$\frac{m_H}{\Delta}(N)$ and 
$N^{-2}$-$\frac{m_H}{\Delta}(N)$ 
fittings,
aiming to appreciate possible systematic errors
other than a fitting error.
Our data in Fig. \ref{figure7}
indicate that the critical amplitude ratio 
$m_H/\Delta$
varies monotonically, as the decay rate $\sigma$
changes.
In other words,
the low-energy spectrum reflects the
long-range-interaction decay rate
$\sigma$
rather sensitively.

We address
a number of remarks.
First, the solution for
the amplitude ratio
$m_H/\Delta$ does not exist for
exceedingly large $\sigma \gtrapprox 1.6$.
That is,
the Higgs particle becomes unstable for
large $\sigma$,
and the plateau as in Fig. \ref{figure6} disappears
(shrinks) eventually.
The end point $\sigma \approx 1.6$ corresponds to
$D \approx 2.25$
according to the $\sigma \leftrightarrow D$ relation
(\ref{sigma-D_relation});
the dimensionality $D=2.25$ 
is about to reach the lower
critical dimension $D=2$.
Last,
the excitation gaps,
$m_H$ and $\Delta$,
as well as the amplitude ratio $m_H/\Delta$
exhibit smooth variations
throughout
 the intermediate regime $2/3 < \sigma <2$.
As for the Ising counterpart \cite{Homrighausen17,Vanderstraeten18},
an anomaly (a dynamical phase transition)
was observed around $\sigma \approx 1.3$.
Correspondingly,
the critical amplitude ratio
$m_2/m_1$ ($m_{i}$: $i$-th excitation gap)
displays a non-monotonic behavior \cite{Rose16};
actually, a minimum locates
in the midst of the intermediate regime
as to $m_2/m_1$.

\subsection{\label{section2_3}
Comparison with the preceding results
via the
$\sigma \leftrightarrow D$ relation \cite{Angelini14,Defenu17}}

According to the elaborated $\epsilon$-expansion analysis,
the critical amplitude ratio 
is given by 
\begin{equation}
\label{epsilon_expansion}
m_H/\Delta=\sqrt{2}
   + \sqrt{2} \epsilon 
   \left(
   \frac{4 \ln2+3\sqrt{3}\pi}{40}
   -\frac{1}{4}
   \right)
,
\end{equation}
with  $D=4-\epsilon$.
As a reference,
in Fig. \ref{figure7},
we present the 
$\epsilon$-expansion result
(dots).
Here, we made use of the 
$\sigma \leftrightarrow D$ relation \cite{Angelini14,Defenu17}
\begin{equation}
\label{sigma-D_relation}
D=2/\sigma+1 ,
\end{equation}
in order to establish a relationship between
the $\epsilon$-expansion result and ours.
The $\sigma \leftrightarrow D$ relation 
has been pursued extensively,
and more refined formulas were proposed.
The above expression
(\ref{sigma-D_relation})
has an advantage in that it is given
in a closed form.

The overall features of our data and the analytic result 
(\ref{epsilon_expansion})
resemble to each other.
Our simulation result suggests that a slight enhancement 
as to the analytic formula for the large-$\sigma$ regime,
$\sigma \gtrapprox 1$.
Remarkably enough,
the above formula 
(\ref{sigma-D_relation})
up to O$(\epsilon^1)$
may be practically of use
to establish
a relationship between
the power of algebraic decay $\sigma$
and
 the low-lying excitation spectrum.

The Higgs mass $m_H$ may have an uncertainty,
because it has a finite life time;
the reciprocal life time corresponds to
the intrinsic $m_H$-peak width.
According to the $\epsilon$-expansion analysis \cite{Katan15},
the scaled peak width $\delta m_H /\Delta$ should 
also take a universal constant 
\begin{equation}
\label{peak_width}
\frac{\delta m_H}{\Delta}=\frac{\sqrt{2}\pi\epsilon}{40} 
 .
\end{equation}
At $\sigma = 1$ ($D=3$), this formula yields the scaled peak width 
$\delta m_H/\Delta=0.11 \dots $.
Taking 
this intrinsic uncertainty into account,
our result and the analytical formula
(\ref{epsilon_expansion}) do not display 
 substantial disagreement
for the small-$\sigma$ side, $\sigma \lessapprox 1$,
As mentioned in Sec. \ref{section2_2},
the series of data $m_H/\Delta$ terminate around $\sigma \approx 1.6$
($D \approx 2.25$).
Such a feature is supported by the analytic formula 
(\ref{peak_width}) in the sense that
for large $\sigma$
(large $\epsilon$), the Higgs particle becomes obscure nonetheless.



We make a brief overview 
for the fixed-$D=3(=2+1)$-lattice analyses.
By means of the quantum Monte Carlo method,
the estimates, 
$m_H/\Delta=2.1(3)$ 
\cite{Gazit13a,Gazit13b}
and $3.3(8)$ 
\cite{Chen13},
were reported,
whereas the
exact diagonalization method yields 
$2.1(2)$ \cite{Nishiyama15}.
With the functional-renormalization-group method,
the results,
$m_H/\Delta=2.4$ 
\cite{Rancon14}
and $2.2$
\cite{Rose15}, were obtained,
while the $\epsilon$-expansion method
\cite{Katan15}
yields $1.67$.
The present simulation admits
$m_H/\Delta = 1.78(11)$ at $\sigma=1$
($D=3$).
These results seem to be rather scattered.
At present, it is not clear whether
the discrepancy should be attributed to
the above-mentioned intrinsic uncertainty (\ref{peak_width}),
or a systematic one.
Actually, the analyticity for the dynamical scalar susceptibility
$\chi_s(\omega)$ \cite{Katan15}
reveals an appreciable discrepancy between the
$m_H$-pole position and the actual peak position;
from the latter, the quantum Monte Carlo results were read off.
Clearly,
detailed information as to the incoherent part of $\chi_s(\omega)$
would be
 desirable so as to resolve the peak position out of the incoherent background.
To the best of author's knowledge,
neither
the $m_H$-peak shape nor the life time
has  been investigated quantitatively.
At present,
the analytic result (\ref{epsilon_expansion}) and ours would be 
``easier to reconcile with the lower of''   
\cite{Katan15}
those preceding results \cite{Gazit13a,Gazit13b,Chen13,Nishiyama15,Rancon14,Rose15}.

\section{\label{section3}
Summary and discussions}

The quantum $XY$ spin chain (\ref{Hamiltonian})
with the algebraically decaying interactions, $1/r^{1+\sigma}$,
was investigated
by means of the exact diagonalization method.
This method enables us to identify the spectral gaps,
$m_H$ and $\Delta$,
by specifying
the quantum numbers to Eqs. (\ref{Higgs_gap}) and
(\ref{paramagnetic_gap}), respectively.
As a preliminary survey,
we analyzed the finite-size-scaling behavior \cite{Defenu17} for $m_H$,
and as a byproduct, we obtained
the $\sigma$-dependent dynamical critical exponent $z$.
The result $z$ seems to obey
the approximate formula 
(\ref{z_formula}), suggesting that the formula 
may be practically of use for a wide range of $\sigma$;
the discrepancies around both upper and lower critical thresholds,
$\sigma=2$ and $2/3$, respectively, should be attributed to the  notorious
corrections to finite-size scaling 
\cite{Luijten02,Brezin14,Defenu15,Fey16}.

We then turn to the analysis of the critical amplitude ratio
$m_H/\Delta$.
Our data obey
the $\epsilon$-expansion result, 
Eq. 
(\ref{epsilon_expansion}),
for the small-$\sigma$ side, $\sigma \lessapprox 1$,
whereas 
the simulation result displays 
a slight enhancement 
as to the analytical formula
for large $\sigma$.
It has to be mentioned that there should exist an
intrinsic 
uncertainty, Eq. (\ref{peak_width}),
as to $m_H/\Delta$
due to the finite life time for $m_H$.
Taking 
this uncertainty into account,
the overlap between
the analytical formula (\ref{epsilon_expansion})
and the numerical result
do not exhibit substantial disagreement in the small-$\sigma$ side,
$\sigma \lessapprox 1$.
For exceedingly large $\sigma$,
the Higgs particle gets unstable,
and the amplitude ratio $m_H/\Delta$ becomes 
obscure nonetheless.
Note that both gaps $m_H$ and $\Delta$ are
experimentally observable \cite{Endres12}
in proximity to the critical point.
It would be intriguing that the amplitude ratio $m_H/\Delta$
reflects the power of algebraic decay $\sigma$
as a monotonic function;
actually,
 the correlation-length critical exponent
$\nu$ displays a
 non-monotonic behavior \cite{Defenu15}
as to the variation of 
 $\sigma$.


As for the fixed-$D=3(=2+1)$ lattice simulation,
the results, $m_H/\Delta=2.1(3)$ 
\cite{Gazit13a,Gazit13b}
and 
$3.3(8)$
\cite{Chen13}, were reported with the quantum Monte Carlo method,
whereas the exact diagonalization method yields 
$2.1(2)$
\cite{Nishiyama15}.
Through the functional-renormalization-group method,
the estimates, $m_H/\Delta=2.4$
\cite{Rancon14}
and 
$2.2$
\cite{Rose15}, were obtained,
while the $\epsilon$-expansion method yields $1.67$.
The present simulation admits $m_H/\Delta = 1.78(11)$ at $\sigma=1$
($D=3$).
These results seem to be rather scattered.
At present, it is not clear whether the disagreement should be attributed to
the above-mentioned
intrinsic uncertainty or a systematic one.
Actually,
the analyticity for the dynamical scalar susceptibility $\chi_s(\omega)$ \cite{Katan15}
reveals an appreciable discrepancy between
the $m_H$-pole position
and
the actual spectral-peak position;
from the latter,
the quantum Monte Carlo results were read off.
Clearly, detailed information as to the incoherent part
of $\chi_s(\omega)$
would be
desirable so as to resolve $m_H$ out of the background.
This problem is left for the future study.






\bibliographystyle{elsarticle-num}







\end{document}